\documentclass[onecolumn,twoside]{IEEEtran}
\usepackage{times}

\usepackage{amsmath}  
\usepackage{amssymb}  
\usepackage{mathrsfs} 

\usepackage{theorem}  
\usepackage{cite}     
\usepackage{comment}  

\usepackage{upref}
\usepackage{amsfonts}

\usepackage{verbatim}

\usepackage[dvipsnames,usenames]{color}
\usepackage{enumerate}

\newcommand{\be}[1]{\begin{equation}\label{#1}}
\newcommand{\ee}{\end{equation}}

\newcommand{\bc}{\begin{center}}
\newcommand{\ec}{\end{center}}

\newcommand{\eq}{\mbox{$\,=\,$}}

\newcommand{\cC}{{\cal C}}

\renewcommand{\leq}{\leqslant}

\renewcommand{\geq}{\geqslant}

\newcommand{\al}{\alpha}

\newcommand{\lan}{\mbox{$\langle$}}
\newcommand{\ran}{\mbox{$\rangle$}}

\newcommand{\uc}{\underline{c}}

\newcommand{\xor}{\oplus}
\newcommand{\qed}{\hfill$\Box$\\[1ex]}

\newcommand{\Cref}[1]{Co\-rol\-la\-ry\,\ref{#1}}

\theoremstyle{plain} \theorembodyfont{\normalfont\slshape}

\newtheorem{thm}{Theorem$\!$}

\newtheorem{prop}[thm]{Proposition$\!$}

\newtheorem{lem}[thm]{Lemma$\!$}
\newenvironment{lemma}{\begin{lem}\hspace*{-1ex}{\bf.}}{\end{lem}}

\newtheorem{cor}[thm]{Corollary$\!$}

\newtheorem{defi}{Definition}

\newtheorem{cl}{Claim}

\newtheorem{alg}{Algorithm}

\theorembodyfont{\normalfont}

\newtheorem{exam}{Example$\!$}
\newenvironment{example}{\begin{exam}\hspace*{-1ex}{\bf .}}{\end{exam}}

\newtheorem{remrk}{Remark$\!$}

\definecolor{Codecolor}{named}{White}  

\newcommand{\Copen}{\mbox{\{\kern-5.50pt\{}}
\newcommand{\Cclose}{\mbox{\}\kern-5.50pt\}}}
\newcommand{\Cslash}{\mbox{$\backslash\kern-6.02pt\backslash$}}

\begin{document}
\title{An efficient implementation of the Shamir secret sharing scheme}

\author{\large Allyson Hineman and Mario Blaum 
\thanks{Allyson Hineman is with the Department of Mathematical Sciences, State
University of New York at Fredonia, Fredonia, NY 14063, USA (e-mail:
\texttt{ahineman@fredonia.edu}). 

Mario Blaum  is with the IBM
Research Division-Almaden, San Jose, CA\,95120, USA (e-mail:
\texttt{mblaum@hotmail.com}).
}}

\maketitle
\begin{abstract}
The Shamir secret sharing scheme requires a Maximum Distance
Separable (MDS) code, and in its
most common implementation, a Reed-Solomon (RS) code is used.
In this paper, we observe that the encoding procedure can be made
simpler and faster by dropping the MDS condition and specifying the possible
symbols that can be shared. In particular, the process can be made
even faster by using array codes based on XOR operations instead of
RS codes.
\end{abstract}
\begin{IEEEkeywords}
Shamir secret sharing scheme, erasure correcting codes, MDS codes,
Reed-Solomon codes, array codes.
\end{IEEEkeywords}


\section{Introduction}
\label{sec1}
Storage systems are under continuous cyber attacks like ransomware,
which have become endemic. It is extremely important to protect these
systems using the most advanced tools in key
distribution. The Shamir secret sharing scheme~\cite{s}, adapted to the cyber
challenges of the present, is one of
such advanced tools. The secret may consist of a
large file and the pieces of information distributed to the participants may
continuously change in order to enhance security, hence the whole
process requires very fast encoding and decoding algorithms. The
purpose of this paper is to present some methods achieving this goal.

The Shamir secret sharing scheme consists of a secret symbol $D$ that can be
reconstructed by sharing $n-1$ symbols among $n-1$ different
participants. Given $k<n$, the
secret symbol can be reconstructed from any $k$ of the $n-1$ shared
symbols by an interpolation process. However, knowledge of any $k-1$
symbols gives no information about the secret $D$. It was
observed~\cite{msa} that the Shamir scheme is equivalent to
implementing an $[n,k]$ MDS code~\cite{ms} such that the secret is one of the
data symbols (for example, the first symbol, a convenient assumption for
implementation, as we will see in the next section), the
remaining $k-1$ data symbols are random symbols, and the $n-k$
parity symbols are obtained by encoding the $k$ data symbols into the
given MDS code. Then, the
$n-1$ symbols excluding the secret symbol are distributed
among $n-1$ participants. Any $k$ participants can then reconstruct
the secret symbol by performing erasure correction, but less than $k$
participants are unable to do so.

We will present a modification of the Shamir scheme in which the
parity symbols are not assigned to participants, but are known by
everybody. Only the $k-1$ data symbols excluding the secret symbol
are assigned. The method will make the encoding faster, since the
parities will be independent from each other, no linear system needs
to be solved and they can even be computed in parallel. The
decoding will be as fast as the one of the traditional Shamir scheme. The
encoding and decoding can be made even faster by using array codes
based on XOR operations, a feature that has been used in RAID-type 
architectures~\cite{Cor+04}.

The paper is structured as follows: in Section~\ref{sec2}, we
describe the modified Shamir scheme and we discuss its advantages
during encoding. In particular, we illustrate this modified scheme with RS codes. In 
Section~\ref{sec3}, we consider the advantages of using the modified
Shamir scheme of Section~\ref{sec2} with array codes as opposed to RS
codes. In particular, we illustrate the ideas with Generalized
EVENODD codes~\cite{bbv}. In Section~\ref{sec4}, we address other
possibilities, like adapting the modified Shamir scheme to
Generalized Row-Diagonal Parity (GRDP)
codes~\cite{b,f} and identifying cases in which some participants report incorrect
symbols. We divide those cases into two categories: one in which some
participants are traitors and deliberately present the wrong symbol,
and another in which a few errors are involuntary. In the second
case, we propose mitigation by using array codes with local properties.


\section{Modified Shamir secret sharing scheme}
\label{sec2}
Assume that $D_0$ is a secret symbol, there are $k-1$ participants and
we want this secret symbol to be recoverable as long as $k-r$
participants are
present, where $r<k$, but a gathering of at most $k-r-1$ participants
provides no information about $D_0$. In this section, we will assume
that $D_0$ is a symbol in a finite
field $GF(q)$~\cite{ms} (for simplicity, we assume that $q$ is a
power of 2 throughout the paper, although this assumption is not
necessary). Assume that the $k-1$ participants are
each assigned a random symbol, say $D_i\in GF(q)$, $1\leq i\leq k-1$.
Let $\cC$ be a $[k,k-r]$ MDS code over $GF(q)$ (for example, a RS
code), and $H$ an $r\times k$ parity-check matrix of $\cC$. Construct
a new $[k+r,k]$ code $\cC'$ whose parity-check matrix is the $r\times
(k+r)$ (systematic) matrix

\begin{eqnarray}
\label{eq1}
H'&\eq &(H\,|\,I_r),
\end{eqnarray}
where $I_r$ is the $r\times r$ identity
matrix. It is well known that if $r\leq 3$, then code $\cC'$ is
MDS~\cite{ms}, but this is not the case for $r\geq 4$. However, it
does not matter if $\cC'$
is not MDS, the result will be valid for any $r<k$.

In effect, assume that $k-r$ participants are present, say, those
holding 
$$D_{j_0},D_{j_1},\ldots,D_{j_{k-r-1}},\quad {\rm where}\quad 1\leq
j_0<j_1<\cdots <j_{k-r-1}\leq k-1,$$ 
while the symbols
$D_0,D_{i_1},D_{i_2},\ldots,D_{i_{r-1}}$ are missing, where 
$$1\leq i_1<i_2<\cdots <i_{r-1}\leq k-1\quad {\rm and}\quad
\{j_0,j_1,\ldots ,j_{k-r-1}\}\cup \{0,i_1,\ldots ,i_{r-1}\}\eq
\{0,1,\ldots,k-1\}.$$
The missing $r$ symbols can be recovered if the corresponding
$r\times r$ submatrix of
the parity-check matrix $H'$ is invertible. Specifically, assume that
$H'\eq (\uc_0,\uc_1,\ldots,\uc_{k+r-1})$, where $\uc_i$ is column $i$
of $H'$, hence, we have to show that the $r\times r$ submatrix $H_r$ of
$H'$ given by $H_r\eq (\uc_0,\uc_{i_1},\uc_{i_2},\ldots,,\uc_{i_{r-1}})$ is
invertible. Since $i_{r-1}\leq k-1$, by~(\ref{eq1}), this submatrix is also a
submatrix of $H$, and since $H$ is the parity-check matrix of an MDS code,
then $H_r$ is invertible~\cite{ms}. In particular, symbol $D_0$,
which corresponds to the secret, can be recovered. This is impossible
if less than $k-r$ participants are present.

Obtaining the parity symbols using the parity-check matrix $H'$
according to~(\ref{eq1}) is
very simple, since matrix $H'$ is in systematic form. Specifically,
$$(D_k,D_{k+1},\ldots,D_{k+r-1})\eq H (D_0,D_1,\ldots,D_{k-1})^T,$$
a process that is very fast for codes such as RS codes (it is equivalent
to computing $r$ syndromes in a RS code). Encoding in a regular RS
code is a special case of the decoding, thus, it involves solving a
linear system of $r$ equations with $r$ unknowns. This is not the
case for the systematic encoding of code $\cC'$, since the parities are computed
independently and no linear system needs to be solved, they may be
even computed in parallel. The resulting code is not MDS when
$r\geq 4$ and $\cC$ is a RS code, but in our case it does not matter, since the
erasures are in the data and, as we have seen, the system is always solvable. 


\begin{example}
\label{ex1}
{\rm
Consider the finite field $GF(8)$ with primitive polynomial~\cite{ms}
$1+x+x^3$. Let $k\eq 7$ and $r\eq 4$, so, according to the
description above, let $\cC$ be a $[7,3]$ RS code over $GF(8)$ with
parity-check matrix

$$
H\eq\left(
\begin{array}{ccccccc}
1&1&1&1&1&1&1\\
1&\al &\al^2&\al^3&\al^4&\al^5&\al^6\\
1&\al^2 &\al^4&\al^6&\al^8&\al^{10}&\al^{12}\\
1&\al^3 &\al^6&\al^9&\al^{12}&\al^{15}&\al^{18}\\
\end{array}
\right).
$$

Also, according to~(\ref{eq1}), $\cC'$ is the $[11,7]$ code whose parity-check matrix $H'$ is

$$
\left(
\begin{array}{ccccccccccc}
1&1&1&1&1&1&1&1&0&0&0\\
1&\al &\al^2&\al^3&\al^4&\al^5&\al^6&0&1&0&0\\
1&\al^2 &\al^4&\al^6&\al^8&\al^{10}&\al^{12}&0&0&1&0\\
1&\al^3 &\al^6&\al^9&\al^{12}&\al^{15}&\al^{18}&0&0&0&1\\
\end{array}
\right).
$$

Next, assume that the secret is the symbol $D_0\eq\al^2$ and the 6
participants are assigned the symbols $D_1\eq\al^3$, $D_2\eq\al$,
$D_3\eq 1$, $D_4\eq 0$, $D_5\eq\al^6$ and $D_6\eq\al^3$. The first
step is computing the parity symbols as

$$(D_7,D_8,D_9,D_{10})\eq H\left(
\begin{array}{c}
\al^2\\\al^3\\\al \\ 1\\ 0\\ \al^6\\ \al^3
\end{array}
\right)\eq (\al , 0,\al^5,\al^2),
$$
i.e., $D_7\eq\al$, $D_8\eq 0$, $D_9\eq\al^5$ and $D_{10}\eq\al^2$.
The parity symbols $D_7$, $D_8$, $D_9$ and $D_{10}$ are known by all
the participants.

Now, assume that we have $k-r\eq 3$ participants, say, $D_2$,
$D_3$ and $D_5$, who want to compute $D_0$. The parity-check matrix
$H'$ gives the following system of 4 equations with 4 unknowns:

\begin{eqnarray*}
D_0\xor D_1\xor D_4\xor D_6&=&S_0\\
D_0\xor \al D_1\xor \al^4D_4\xor \al^6D_6&=&S_1\\
D_0\xor \al^2D_1\xor \al^8D_4\xor \al^{12}D_6&=&S_2\\
D_0\xor \al^3D_1\xor \al^{12}D_4\xor \al^{18}D_6&=&S_3,
\end{eqnarray*}
where $S_0\eq D_2\xor D_3\xor D_5\xor D_7\eq\al^2$, $S_1\eq \al^2D_2\xor
\al^3D_3\xor\al^5D_5\xor D_8\eq \al^4$, $S_2\eq \al^4D_2\xor
\al^6D_3\xor \al^{10}D_5\xor D_9\eq 1$ and $S_3\eq \al^6D_2\xor
\al^9D_3\xor \al^{15}D_5\xor D_{10}\eq 0$.

We need to solve the system above only for the secret symbol $D_0$.
For example, using Cramer's rule, we have

\begin{eqnarray*}
D_0&=&\frac{
\det\left(
\begin{array}{cccc}
\al^2&1&1&1\\
\al^4&\al & \al^4&\al^6\\
1&\al^2 & \al^8&\al^{12}\\
0&\al^3 & \al^{12}&\al^{18}
\end{array}
\right)
}{\det\left(
\begin{array}{cccc}
1&1&1&1\\
1&\al & \al^4&\al^6\\
1&\al^2 & \al^8&\al^{12}\\
1&\al^3 & \al^{12}&\al^{18}
\end{array}
\right)}\;\eq\; \al^2,
\end{eqnarray*}
since the determinant of the numerator is $\al$, while the
determinant of the denominator is a Vandermonde determinant, which is
equal to $$(1\xor \al) (1\xor \al^4)(1\xor \al^6)(\al\xor
\al^4)(\al\xor \al^6)(\al^4\xor \al^6)\eq \al^6.$$

We will see next a more efficient method for computing the erased
symbol $D_0$.
\qed
}
\end{example}

Example~\ref{ex1} illustrates the simplicity of the encoding method: each parity symbol
is the syndrome of the $k$ data symbols with respect to the
parity-check matrix $H$. For RS codes, there is ample literature
on how to efficiently compute the syndromes. Regarding the decoding,
mainly the computation of the secret symbol $D_0$, we will next describe
a method that is similar to the one presented in~\cite{br}.

In effect, assume the conditions described above with the codes $\cC$
and $\cC'$, where the $r$ erased symbols are\\
$D_0,D_{i_1},D_{i_2},\ldots,D_{i_{r-1}}$, the symbols
corresponding to the $k-r$ participants that are present are
$D_{j_0},D_{j_1},\ldots,D_{j_{k-r-1}}$, while the parity symbols are
$D_{k},D_{k+1},\ldots,D_{k+r-1}$. Moreover, we assume that $\cC$ is a
(shortened) RS code with parity-check matrix

\begin{eqnarray}
\label{eq2}
H\eq \left(
\begin{array}{ccccc}
1&1&1&\ldots &1\\
1&\al &\al^2&\ldots &\al^{k-1}\\
\vdots&\vdots&\vdots&\ddots&\vdots\\
1&\al^{r-1} &\al^{2(r-1)}&\ldots &\al^{(k-1)(r-1)}
\end{array}
\right).
\end{eqnarray}

We are interested in
computing only $D_0$. The syndrome $S_u$, $0\leq u\leq r-1$, is given by

\begin{eqnarray}
\nonumber
S_u&=&
\left(\bigoplus_{v=0}^{k-r-1}\al^{uj_v}D_{j_v}\right)\xor D_{k+u}\\
\label{eq3}
&=&\bigoplus_{s=0}^{r-1}\al^{ui_s}D_{i_s}.
\end{eqnarray}

Define the polynomials of degree at most $r-1$
\begin{eqnarray}
\label{eq4}
S(x)&=& S_0\xor
S_1x\xor\cdots\xor S_{r-1}x^{r-1}
\end{eqnarray}
and

\begin{eqnarray}
\nonumber
G(x)&=&(x\xor\al^{i_1})(x\xor\al^{i_2})\cdots (x\xor\al^{i_{r-1}})\\
\label{eq5}
&=&g_{r-1}\xor g_{r-2}x\xor\cdots\xor g_{0}x^{r-1}.
\end{eqnarray}

Notice that $G(1)\eq(1\xor\al^{i_1})(1\xor\al^{i_2})\cdots
(1\xor\al^{i_{r-1}})$ while $G(\al^{i_s})\eq 0$ for $1\leq s\leq r-1$.
Then, assuming $i_0\eq 0$, by~(\ref{eq3}), (\ref{eq4}) and~(\ref{eq5}), we have

\begin{eqnarray}
\nonumber
\bigoplus_{u=0}^{r-1}S_ug_{r-u-1}&=&\bigoplus_{u=0}^{r-1}\left(
\bigoplus_{s=0}^{r-1}\al^{ui_s}D_{i_s}\right)g_{r-u-1}\\
\nonumber
&=&\bigoplus_{s=0}^{r-1}D_{i_s}\left(\bigoplus_{u=0}^{r-1}
g_{r-u-1}\al^{ui_s}\right)\\
\nonumber
&=&\bigoplus_{s=0}^{r-1}D_{i_s}G(\al^{i_s})\\
\label{eq6}
&=&D_{0}\prod_{s=1}^{r-1}(1\xor\al^{i_s}).
\end{eqnarray}

Thus, by~(\ref{eq5}) and~(\ref{eq6})

\begin{eqnarray}
\label{eq7}
D_{0}&=&\frac{\bigoplus_{u=0}^{r-1}S_ug_{r-u-1}}{\prod_{s=1}^{r-1}(1\xor\al^{i_s})}
\eq \frac{\bigoplus_{u=0}^{r-1}S_ug_{r-u-1}}{\bigoplus_{u=0}^{r-1}g_u},
\end{eqnarray}
since by~(\ref{eq5}), $G(1)\eq\prod_{s=1}^{r-1}(1\xor\al^{i_s})\eq\bigoplus_{u=0}^{r-1}g_u$.
Both the numerator and the denominator in~(\ref{eq7})
can be easily computed.

\begin{example}
\label{ex2}
{\rm
Let us revisit Example~\ref{ex1} and find $D_0$ using~(\ref{eq7}).
Using the syndromes obtained in Example~\ref{ex1} and~(\ref{eq4}), we
obtain

\begin{eqnarray*}
S(x)&=& \al^2\xor\al^4x\xor x^2.
\end{eqnarray*}

From Example~\ref{ex1}, we have $i_1\eq 1$, $i_2\eq 4$ and $i_3\eq
6$, so, by~(\ref{eq5}), we obtain

\begin{eqnarray*}
G(x)&=& (x\xor\al )(x\xor\al^4)(x\xor\al^6)\eq\al^4\xor\al^6x\xor x^2\xor x^3,
\end{eqnarray*}
i.e., $g_3\eq\al^4$, $g_2\eq\al^6$, $g_1\eq 1$ and $g_0\eq 1$
in~(\ref{eq5}). Hence, the numerator in~(\ref{eq7}) is given by

\begin{eqnarray*}
g_3S_0\xor g_2S_1\xor g_1S_2\xor g_0S_3\eq \al^6\xor\al^3\xor 1\eq\al^5,
\end{eqnarray*}
while the denominator equals

\begin{eqnarray*}
G(1)&=&\al^4\xor\al^6\xor 1\xor 1\eq \al^3,
\end{eqnarray*}
so $D_0\eq \al^2$, which coincides with the value obtained in Example~\ref{ex1}.
\qed
}
\end{example}

\section{Use of array codes in the modified Shamir secret sharing scheme}
\label{sec3}
The purpose of using array codes in RAID-type 
architectures~\cite{bbbm,{Cor+04}} was to replace finite field operations, which
usually require a look-up table, by XOR operations. In an application
like the Shamir scheme described in Section~\ref{sec2}, if the size
of the secret is pretty large, implementation of a RS
code has to be done multiple times. Array codes like the ones
described in~\cite{b,bbbm,bbv,br,Cor+04,f} can have symbols (which
correspond to columns in the array) of size
$p-1$, where $p$ is a prime number. Certainly, $p$ can be as large as
needed, while large symbols in a RS code require a large look-up
table in the corresponding finite field and may not be practical.

An example of an MDS array code is given by Blaum-Roth (BR)
codes~\cite{br}. We are not the first to point out the
usefulness of array codes in the context of the Shamir secret sharing
scheme. For example, in~\cite{wd}, the use of BR codes is proposed.

Given an odd prime number $p$, the codewords of a
$[p,k]$ BR code consist of $(p-1)\times p$ arrays such that,
when appending a zero row to such an array in the code, making it a
$p\times p$ array, the lines of slope $i$ (with a toroidal topology),
$0\leq i\leq p-k-1$, have even parity. For example, the first four
lines of the $5\times 5$
array below are in a $[5,2]$ BR code: the horizontal lines (slope 0), the lines
of slope 1 and the lines of slope 2 have all even parity. In the left
array, we illustrate in bold the second line of slope 1, while in the
right array, in bold is the third line of slope 2 (we assume that the
individual symbols in the arrays are bits, although they can have any
size. It is not necessary either that the number of columns is a
prime number, since some columns may be assumed to be zero).

$$\begin{array}{cc}
\begin{array}{|c|c|c|c|c|}
\hline
1& {\bf 0}& 1& 1& 1\\
\hline
{\bf 0}& 0& 0& 1& 1\\
\hline
1& 0& 1& 0& {\bf 0}\\
\hline
1& 1& 0& {\bf 0}& 0\\
\hline\hline
0& 0& {\bf 0}& 0& 0\\
\hline
\end{array}
&
\begin{array}{|c|c|c|c|c|}
\hline
1& {\bf 0}& 1& 1& 1\\
\hline
0& 0& 0& {\bf 1}& 1\\
\hline
{\bf 1}& 0& 1& 0& 0\\
\hline
1& 1& {\bf 0}& 0& 0\\
\hline\hline
0& 0& 0& 0& {\bf 0}\\
\hline
\end{array}
\end{array}
$$

An equivalent algebraic definition of BR codes (and a very convenient
one for decoding) is that they are RS codes over the ring of polynomials modulo
$M_p(x)\eq 1\xor x\xor x^2\xor\cdots\xor x^{p-1}$~\cite{br}. The
parity-check matrix $H$ of such a code (shortened to $k$ columns) is given by~(\ref{eq2}). Let us
point out that the polynomial $M_p(x)$ may not be irreducible (for
example, $M_5(x)$ is irreducible but $M_7(x)\eq (1\xor x\xor
x^3)(1\xor x^2\xor x^3)$). However, the code is always MDS~\cite{br}.

In order to apply our particular version of the Shamir scheme as
described in Section~\ref{sec2}, we need to consider the parity-check
matrix $H'$ as given by~(\ref{eq1}), while $H$ is given
by~(\ref{eq2}). Such a resulting code is a generalization of the
EVENODD code~\cite{bbbm} and has different names in literature:
generalized EVENODD code~\cite{bbv}, independent parity (IP)
code~{\cite{bbv}} or Blaum-Bruck-Vardy code~\cite{hsl}. The MDS
condition of these codes has been extensively studied for $r\geq
4$~\cite{bbv,hsl}, but for our purpose the modified Shamir scheme
will always work for $r<k$, as in the case of RS codes we studied in
Section~\ref{sec2}. Notice that for these generalized EVENODD codes,
the horizontal lines always have even parity, while the
lines of slope $i$, $1\leq i\leq r-1$, may have either even or odd
parity: the special line of slope $i$ starting in the last bit of the
first column (which is 0 and not written) determines the parity of
all the other lines of slope $i$~\cite{bbv}. So, the encoding is very
fast and convenient.

\begin{example}
\label{ex3}
{\rm
The following array corresponds to a generalized EVENODD code with
$p\eq 5$ and 3 parities:

$$\begin{array}{cc}
\begin{array}{|c|c|c|c|c||c|c|c|}
\hline
1& 0& 0& 0&\hspace{-.1cm} {\bf 0}\hspace{-.1cm}& 1& 0& 0\\
\hline
0& 0& 1&\hspace{-.1cm} {\bf 1}\hspace{-.1cm}& 0& 0& 1& 0\\
\hline
1& 0& \hspace{-.1cm}{\bf 0}\hspace{-.1cm}& 1& 0& 0& 0& 1\\
\hline
1&\hspace{-.1cm} {\bf 1}\hspace{-.1cm}& 0& 1& 1& 0& 0& 1\\
\hline\hline
\hspace{-.1cm}{\bf 0}\hspace{-.1cm}& 0& 0& 0& 0& 0& 0& 0\\
\hline
\end{array}
&
\begin{array}{|c|c|c|c|c||c|c|c|}
\hline
1& 0& \hspace{-.1cm} {\bf 0}\hspace{-.1cm}& 0& 0& 1& 0& 0\\
\hline
0& 0& 1& 1& \hspace{-.1cm} {\bf 0}\hspace{-.1cm}& 0& 1& 0\\
\hline
1& \hspace{-.1cm} {\bf 0}\hspace{-.1cm}& 0& 1& 0& 0& 0& 1\\
\hline
1& 1& 0& \hspace{-.1cm} {\bf 1}\hspace{-.1cm}& 1& 0& 0& 1\\
\hline\hline
\hspace{-.1cm} {\bf 0}\hspace{-.1cm}& 0& 0& 0& 0& 0& 0& 0\\
\hline
\end{array}
\end{array}
$$

In the array on the left we illustrate in bold the entries of the
special line of slope 1 starting at the bottom of the first column.
It has an even number of ones, so all the diagonals must have even
parity, which is determined by the second parity column (the first
parity column corresponds to horizontal parity, so it has always even
parity). Similarly, in the array on the right, we illustrate in bold 
the entries corresponding to the special line of slope 2 starting at
the bottom of the first column. In this case, the number of 1s of
this special line is odd, so all the lines of slope 2 must have odd
parity, and this is reflected in the last parity column. Notice that
the parities are independent of each other, so, for that reason,
these codes are also called Independent Parity (IP) codes~\cite{bbv}.

Denote the 8 columns in the array as $(\uc_0,\uc_1,\ldots,\uc_7)$ and
assume that the secret is $\uc_0$, while the parities are $\uc_5$,
$\uc_6$ and $\uc_7$. The four data columns $\uc_1$,
$\uc_2$, $\uc_3$ and $\uc_4$ are assigned to participants, while the
three parity columns are known by everybody. Assume that $k-r\eq 2$
participants get
together, say, $\uc_2$ and $\uc_4$. Then, symbols $\uc_0$ (the
secret), $\uc_1$ and $\uc_3$ are erased, and we have to use $\uc_2$,
$\uc_4$, $\uc_5$, $\uc_6$ and $\uc_7$ to retrieve them. We proceed
similarly to the method described in Section~\ref{sec2} for RS codes.
The first step is computing the syndromes using the parity-check matrix
$H'$:

\begin{eqnarray*}
S_0&=&\uc_2\xor\uc_4\xor\uc_5\\
S_1&=&\al^2\uc_2\xor\al^4\uc_4\xor\uc_6\\\
S_2&=&\al^4\uc_2\xor\al^8\uc_4\xor\uc_7.
\end{eqnarray*}

Notice that as a function of $\al$, from the array above, $\uc_2\eq
\al$, $\uc_4\eq \al^3$, $\uc_5\eq 1$, $\uc_6\eq \al$ and
$\uc_7\eq\al^2\xor\al^3$, where $M_5(\al)\eq 0$. Hence, $\al^4\eq
1\xor\al\xor\al^2\xor\al^3$ and $\al^5\eq 1$, and the syndromes can
be easily calculated as $S_0\eq 1\xor\al\xor\al^3$, $S_1\eq
\al\xor\al^2\xor\al^3$ and $S_2\eq 1\xor\al\xor\al^2\xor\al^3\eq\al^4$. Thus,
by~(\ref{eq4}),

\begin{eqnarray*}
S(x)&=&(1\xor\al\xor\al^3)\xor (\al\xor\al^2\xor\al^3)x\xor \al^4x^2.
\end{eqnarray*}

Next, using~(\ref{eq5}), since $i_1\eq 1$ and $i_2\eq 3$

\begin{eqnarray*}
G(x)&=&(x\xor\al)(x\xor\al^3)\\
&=&\al^4\xor (\al\xor\al^3)x\xor x^2,
\end{eqnarray*}
i.e., in~(\ref{eq5}), $g_2\eq \al^4$,
$g_1\eq \al\xor\al^3$ and $g_0\eq 1$. Next, we have to compute the
right hand side in~(\ref{eq6}), which gives

\begin{eqnarray*}
S_0g_2\xor S_1g_1\xor S_2g_0&=& \al\xor\al^3.
\end{eqnarray*}

Using~(\ref{eq6}), we have to solve

\begin{eqnarray*}
(1\xor\al)(1\xor\al^3)D_0&=& \al\xor\al^3,
\end{eqnarray*}

Let $(1\xor\al^3)D_0\eq X$, then we have to solve first

\begin{eqnarray}
\label{eq72}
(1\xor\al)X&=& \al\xor\al^3,
\end{eqnarray}
which can be done using the following lemma~\cite{br}:

\begin{lemma}
\label{l1}
{\rm
Assume that we want to solve $(1\xor\al^j)X(\al)\eq
Y(\al)$ over the ring of polynomials modulo $M_p(x)$, where $p$ is
prime, $1\leq j\leq p-1$, 
$Y(\al)\eq \bigoplus_{i=0}^{p-2}y_i\al^i$ is given and
$X(\al)\eq \bigoplus_{i=0}^{p-2}x_i\al^i$. Then, for $1\leq u\leq p-1$,

\begin{eqnarray}
\label{eq71}
x_{\lan -uj-1\ran}&=&x_{\lan -(u-1)j-1\ran}\xor\hat{y}_{\lan
-(u-1)j-1\ran},
\end{eqnarray}
where given any integer $m$, $\lan m\ran$ denotes the unique integer
$v$, $0\leq v\leq p-1$, such 
that $v\equiv m\;(\bmod\;p)$ (for example, for $p=5$, $\lan -2\ran=3$),
$y_{p-1}\eq 0$ and $\hat{y}_j\eq y_j\xor\bigoplus_{i=0}^{p-1}y_i$.
\qed }
\end{lemma}

Applying recursion~(\ref{eq71}) in Lemma~\ref{l1} to~(\ref{eq72}), we obtain
$X\eq \al\xor\al^2$. Next we have solve $(1\xor\al^3)D_0\eq \al\xor\al^2.$
Again, 
using recursion~(\ref{eq71}), we obtain $D_0\eq
1\xor\al^2\xor\al^3$, which coincides with column $\uc_0$ of the
array above, corresponding to the secret.
\qed
}
\end{example}

\section{Other possibilities and conclusions}
\label{sec4}
The Shamir scheme can have other implementations as well. Another
array code that can be used is the GRDP
code~\cite{b,Cor+04,f}. The GRDP code has a minimal
number of encoding operations, so it is very convenient for the
modified Shamir scheme described in sections~\ref{sec2}
and~\ref{sec3}. 
A $GRDP(p,r)$ code consists of the arrays
$(a_{i,j})_{\substack{0\leq i\leq p-1 \\ 0\leq j\leq p+r-1}}$, such that

\begin{eqnarray}
\label{eq8}
a_{i,p-1}&=&\bigoplus_{j=0}^{p-2}a_{i,j}\\
\label{eq9}
a_{i,k+u}&=&\bigoplus_{j=0}^{p-1}a_{\lan i-uj\ran,j}\quad {\rm
for}\quad 1\leq u\leq r-1,
\end{eqnarray}
where $\lan m\ran$ was defined in Lemma~\ref{l1}.
For example, according to~(\ref{eq8}) and~(\ref{eq9}), the following
is an array in $GRDP(5,3)$:

$$
\begin{array}{|c|c|c|c|c|c|c|}
\hline
1&1&1&1&0&1&0\\\hline
1&0&0&1&0&1&0\\\hline
0&0&1&0&1&0&0\\\hline
1&0&0&0&1&0&1\\\hline\hline
0&0&0&0&0&0&0\\\hline
\end{array}
$$

Column $p+u$, $1\leq u\leq r-1$, contains the parity of the lines of
slope $u$ in the
$p\times p$ array, computed using the horizontal parity (hence, the
parities are not independent as in the extended EVENODD code
described in Section~\ref{sec3}), and excluding
the line starting at location $p-1$ of the first column. The encoding
is simpler than the encoding of the extended EVENODD codes,
since the lines of slope $u$, $1\leq u\leq r-1$, all have even parity
and the parity of the lines starting at location $p-1$ of the first column do
not need to be computed.

From the above discussion, the modified Shamir scheme with GRDP codes is as
follows: let the secret be a symbol
of length $p-1$, where $p$ is prime, then take $p-2$ random symbols
of length $p-1$, and
encode these $p-1$ symbols into a $GRDP(p,r)$ code. The
$p-2$ random symbols together with the horizontal parity symbol are
distributed to $p-1$ participants, while the $r-1$ parity symbols
corresponding to lines of slope $u$, $1\leq u\leq r-1$, are known by
all the participants. Then, if any $p-r$ participants share
their symbols, $r$ erasures can be corrected by the code.

It has been shown~\cite{b,hhsl} that a GRDP code is MDS if
and only if a corresponding generalized EVENODD code is also MDS.
This property
also helps with the decoding in the recovery of the secret: once the
transformation is established, there are efficient methods to decode
the generalized EVENODD code~\cite{b,hhsl} that can be used in our context.

As pointed out in~\cite{msa}, since an MDS code can correct errors
together with erasures, the Shamir scheme can handle cases in which a
number of participants, for a variety of reasons, incorrectly report
their symbols. Specifically, an $[n,k]$ MDS code can correct $s$
errors together with $t$ erasures as long as $2s+t\leq
n-k$~\cite{ms}. In the Shamir scheme, this means that if $n-t$
participants get together and $s$ of them report the wrong symbol,
then the secret can be recovered as long as $2s+t\leq n-k$.

This scheme works also for our modified Shamir scheme: in this case
the $r$ parity symbols are known by everybody, the secret is the
first symbol and the remaining $k-1$ symbols are distributed among
participants. If $k-t$ participants share their symbols but $s$ of
them provide the wrong symbol, then the secret can be recovered as
long as $2s+t\leq r$.

The decoding of RS codes containing both errors and erasures is well
known. However, there is no known efficient decoding algorithm
correcting more than three errors for array codes such as BR codes.
For example, an
efficient algorithm correcting one error and any number of erasures
was presented in~\cite{br}. Efficient algorithms correcting two and
three errors with any number of erasures can be found in~\cite{bbv}.
Beyond that, the problem is open, though
correction of up to three errors may be enough for most
applications of the Shamir scheme.

The inaccuracy of sharing symbols with other participants, as stated
above, may be due to a few different reasons. One such cause involves
a traitor among the participants, who may exploit the information
from the other participants either to have sole access to the secret or to
sabotage the entire enterprise.
Provided that there is enough redundancy, the scheme for correcting
errors and erasures prevents this scenario, allowing for the
identification of up to $s$ traitors.
However, such a scheme is costly if the
participant providing erroneous information did not 
have nefarious purposes. The information may have been
corrupted by a few erroneous or erased bits through normal
noise during transmission of the symbol.

Recently, an expansion of the BR, generalized EVENODD and
GRDP codes was presented~\cite{bh}. In these expansions, the
arrays have column size $p$ as opposed to $p-1$. The expanded codes continue
to be MDS, but each column is in a cyclic code with generator
polynomial $(1\xor x)g(x)$, where $g(x)$ divides $1\xor x^p$. If the
cyclic code has minimum distance $d$, then $s$ bits in error together with
$t$ erased bits can be corrected in every column as long as $2s+t\leq
d-1$. Hence, a few errors and erasures can be corrected locally
in each column of the array without invoking the other columns. The
full power of the code is reserved for cases in which traitors
deliberately misrepresent the column they had been assigned. A
further generalization was obtained in~\cite{wh}, which describes
a generalization of the expanded BR codes to powers of prime numbers.

We presented the decoding algorithm to obtain the secret as a
result of repeated recursions. There are more efficient decoding
algorithms reducing the number of recursions when obtaining all the
erasures, mainly through the LU factorization of Vandermonde
matrices~\cite{wh}. For our purpose, however, we only need to obtain one
erasure, the one corresponding to the secret.

The modified Shamir secret sharing scheme presented in this paper
consists of assigning $k-1$ random data symbols to participants (excluding
the secret), while the parity symbols are independent from each other
and known by everyone.
This method simplifies the encoding since computing the parity does
not require solving a system of linear equations and can be done in
parallel, while the decoding remains the same. 
We studied this modified scheme with RS and with
array codes. By using array codes with local properties, we showed
that the cases in which participants report their
symbols with involuntary errors can be mitigated.

\end{document}